\documentclass[letterpaper, 12pt]{article}[2025/05/19]
\usepackage[english]{babel}
\usepackage{amsfonts,amsmath,amssymb,amsthm,latexsym,amscd,mathrsfs,bbm}
\usepackage{ifthen,cite}
\usepackage[bookmarksnumbered=true]{hyperref}
\usepackage{times,fancyhdr}

\usepackage{amsmath}
\numberwithin{equation}{section}

\usepackage{graphicx}

\hypersetup{pdfpagetransition={Split}}

\newcommand{\evenhead}{Author \ name}
\newcommand{\oddhead}{Article \ name}
\newcommand{\theArticleName}{Article \ name}

\newcommand{\FirstPageHeading}[1]{\thispagestyle{empty}%
\noindent\raisebox{0pt}[0pt][0pt]{\makebox[\textwidth]{\protect\footnotesize \sf }}\par}

\newcommand{\ArticleName}[1]{\renewcommand{\theArticleName}{#1}\vspace{-2mm}\par\noindent {\LARGE\bf  #1\par}}
\newcommand{\Author}[1]{\vspace{5mm}\par\noindent {\Large  #1\par} \par\vspace{2mm}\par}
\newcommand{\Address}[1]{\vspace{2mm}\par\noindent {\it #1} \par}
\newcommand{\Email}[1]{\ifthenelse{\equal{#1}{}}{}{\par\noindent {\rm E-mail: }{\it  #1} \par}}
\newcommand{\URLaddress}[1]{\ifthenelse{\equal{#1}{}}{}{\par\noindent {\rm URL: }{\tt  #1} \par}}
\newcommand{\EmailD}[1]{\ifthenelse{\equal{#1}{}}{}{\par\noindent {$\phantom{\dag}$~\rm E-mail: }{\it  #1} \par}}
\newcommand{\URLaddressD}[1]{\ifthenelse{\equal{#1}{}}{}{\par\noindent {$\phantom{\dag}$~\rm URL: }{\tt  #1} \par}}

\newcommand{\Abstract}[1]{\vspace{6mm}\par\noindent\hspace*{10mm}
\parbox{140mm}{\small {\bf Abstract.} #1}\par}
\newcommand{\Keywords}[1]{\vspace{3mm}\par\noindent\hspace*{10mm}
\parbox{140mm}{\small {\bf Key words:} \rm #1}\par}
\newcommand{\Classification}[1]{\vspace{3mm}\par\noindent\hspace*{10mm}
\parbox{140mm}{\small {\it 2020 Mathematics Subject Classification:} \rm #1}\vspace{3mm}\par}
\newcommand{\ShortArticleName}[1]{\renewcommand{\oddhead}{#1}}
\newcommand{\AuthorNameForHeading}[1]{\renewcommand{\evenhead}{#1}}

\long\def\@makecaption#1#2{
  \sbox\@tempboxa{\small \textbf{#1.}\ \ #2}%
  \ifdim \wd\@tempboxa >\hsize
    {\small \textbf{#1.}\ \ #2}\par \else
    \global \@minipagefalse
    \hb@xt@\hsize{\hfil\box\@tempboxa\hfil}%
  \fi \vskip\belowcaptionskip}

\def\numberwithin#1#2{\@ifundefined{c@#1}{\@nocounterr{#1}}{%
  \@ifundefined{c@#2}{\@nocnterr{#2}}{%
  \@addtoreset{#1}{#2}%
  \toks@\@xp\@xp\@xp{\csname the#1\endcsname}%
  \@xp\xdef\csname the#1\endcsname
    {\@xp\@nx\csname the#2\endcsname.\the\toks@}}}}
\def\E^#1{{\buildrel #1 \over\vee}}
\newtheorem{theorem}{Theorem}

{\theoremstyle{definition}

}

\usepackage{tikz}

\begin{document}

\FirstPageHeading{V.I. Gerasimenko}

\ShortArticleName{Cumulants of Operator Semigroups}

\AuthorNameForHeading{V.I. Gerasimenko}

\ArticleName{\textcolor{blue!50!black}{Cumulants of Operator Semigroups \\ for Hierarchies of Many-Particle Evolution Equations}
}

\Author{V.I. Gerasimenko$^\ast$\footnote{E-mail: \emph{gerasym@imath.kyiv.ua}}}

\Address{$^\ast$ Institute of Mathematics of the NAS of Ukra\"{\i}ne,\\
    \hspace*{4mm}3, Tereshchenkivs'ka Str.,\\
    \hspace*{4mm}01601, Ky\"{\i}v-4, Ukra\"{\i}ne}

\bigskip

\Abstract{The paper presents a brief overview of recent advances in the mathematical theory
of evolution equations for quantum many-particle systems.

The structure of expansions representing non-perturbative solutions to the Cauchy problem
for hierarchies of fundamental evolution equations for many quantum particles is outlined.
It has been shown that the concept of cumulants of operator semigroups forms the basis for
non-perturbative representation of solutions to hierarchies of evolution equations that
describe the state and observables.

It is proven that the constructed dynamics of correlations underlie the description of the
evolution of the state, both for finitely and infinitely many particles, governed by the
BBGKY hierarchy for reduced density operators or by the hierarchy of nonlinear evolution
equations for reduced correlation operators.
In fact, it is demonstrated that the behavior of particle correlations governs the dynamics
of states in quantum systems with an infinite number of particles.
}

\bigskip

\Classification{35C10, 35Q20, 82C05, 82D05, 70F45}

\makeatletter
\renewcommand{\@evenhead}{
\hspace*{-3pt}\raisebox{-7pt}[\headheight][0pt]{\vbox{\hbox to \textwidth {\thepage \hfil \evenhead}\vskip4pt \hrule}}}
\renewcommand{\@oddhead}{
\hspace*{-3pt}\raisebox{-7pt}[\headheight][0pt]{\vbox{\hbox to \textwidth {\oddhead \hfil \thepage}\vskip4pt\hrule}}}
\renewcommand{\@evenfoot}{}
\renewcommand{\@oddfoot}{}
\makeatother

\newpage
\vphantom{math}

\protect\textcolor{blue!50!black}{\tableofcontents}

\vspace{0.7cm}

\textcolor{blue!55!black}{\section{Introduction}}\label{sec1}

Recently, significant progress has been made in mathematical studies describing the evolution of quantum
many-particle systems. This progress was largely driven by contemporary challenges to the theory of quantum
information transfer processes, particularly the description of the evolution of state correlations within
complex systems \cite{MW,SMLR24,FKMS}, and by the problem of deriving quantum kinetic equations as the scaling
asymptotics of solutions to hierarchies of fundamental evolution equations \cite{BogLect,BPS,P2021}. In modern
research, the primary method for deriving kinetic equations is to construct appropriate scaling asymptotics for
a perturbative solution to the Cauchy problem for the BBGKY (Bogolyubov--Born--Green--Kirkwood--Yvon) hierarchy
\cite{BPS,P2021,HP23,Ann23,BGSRS22,CGP97,GF}.

This article aims to formulate and explore the concept of cluster and cumulant expansions of operator groups
(operator semigroups) that underlie the description of the evolution of states and observables of quantum
many-particle systems. The generating operators for the non-perturbative representation of solutions of the
hierarchies of evolution equations for a state of many-particle systems, as well as of the hierarchy of evolution
equations for observables, are determined by the cumulants (semi-invariants) of the dual operator groups and by
the cumulants of operator groups, respectively.

The main sections (Sections 2 and 3) of the article discuss the hierarchies of evolution equations that describe
the evolution of a state of a system consisting of many quantum particles, and provide a classification of these
equations. To this end, in Section 2, we introduce the concept of cumulants (semi-invariants) of operator semigroups.
The appendix examines an alternative approach to describing evolution in terms of observables of systems of quantum
particles.

\textcolor{blue!55!black}{\section{Cluster and cumulant expansions of operator semigroups}}\label{sec2}

To determine the generating operators for the expansions of solutions to the Cauchy problem related to hierarchies
of evolution equations for quantum systems with many particles, we will introduce groups of operators associated
with quantum systems of a finite number of particles, along with their cumulant expansions.

\textcolor{blue!55!black}{\subsection{Preliminaries: Operator groups of the Heisenberg and von Neumann equations}}

In the following, we consider quantum systems of a non-fixed number of identical spinless particles, i.e., an arbitrary
but finite average number of particles that obey the Maxwell--Boltzmann statistics, and use a system of units where
$h={2\pi\hbar}=1$ is Planck's constant and $m =1$ is the particle mass. Let $\mathcal{H}_n=\mathcal{H}^{\otimes n}$ be
an $n$-particle Hilbert space. The Fock space over the space $\mathcal{H}$ is denoted by
$\mathcal{F}_{\mathcal{H}}={\bigoplus\limits}_{n=0}^{\infty}\mathcal{H}_{n}$.

Let $\mathfrak{L}(\mathcal{H}_{n})$ be the space of bounded self-adjoint operators $b_{n}\in\mathfrak{L}(\mathcal{H}_{n})$
with operator norm $\|.\|_{\mathfrak{L}(\mathcal{H}_{n})}$. The self-adjoint operator $b_{n}$ is defined on the space
$\mathcal{H}_{n}=\mathcal{H}^{\otimes n}$, hereinafter also denoted by the symbol $b_{n}(1,\ldots,n)$, which for arbitrary
values of the indices $(i_{1},\ldots,i_{n})\in(1,\ldots,n)$ satisfies the symmetry condition:
$b_{n}(1,\ldots,n)=b_{n}(i_{1},\ldots,i_{n})$.
Let us introduce the space $\mathfrak{L}_\gamma(\mathcal{F}_{\mathcal{H}})$ of sequences of bounded operators
$b=(b_0,b_1(1),\ldots,b_n(1,\ldots,n),\ldots)$ with the norm
\begin{eqnarray*}
   &&\|b\|_{\mathfrak{L}_\gamma(\mathcal{F}_{\mathcal{H}})}=
      \max_{n\geq0}\frac{\gamma^n}{n!}\,\|b_{n}\|_{\mathfrak{L}(\mathcal{H}_{n})},
\end{eqnarray*}
where $0<\gamma<1$ is a real number. The subspace of finite sequences of degenerate operators whose kernels are
infinitely differentiable functions with compact supports is denoted by $\mathfrak{L}_0(\mathcal{F}_{\mathcal{H}})$.

Accordingly, let $\mathfrak{L}^{1}(\mathcal{H}_{n})$ be the space of trace-class operators
$f_{n}\equiv f_{n}(1,\ldots,n)\in\mathfrak{L}^{1}(\mathcal{H}_{n})$, which for arbitrary values of the indices
$(i_{1},\ldots,i_{n})\in(1,\ldots,n)$ satisfy the symmetry condition: $f_{n}(1,\ldots,n)=f_{n}(i_{1},\ldots,i_{n})$,
with the norm: $\|f_{n}\|_{\mathfrak{L}^{1}(\mathcal{H}_{n})}=\mathrm{Tr}_{1,\ldots,n}|f_{n}(1,\ldots,n)|,$ where the
symbol $\mathrm{Tr}_{1,\ldots,n}$ denote the partial traces of the operator $f_{n}$. Let us introduce the space
$\mathfrak{L}^{1}_\alpha(\mathcal{F}_{\mathcal{H}})$ of sequences of trace-class operators
$f=(f_0,f_1(1),\ldots,f_n(1,\ldots,n),\ldots)$ with the norm
\begin{eqnarray*}
   &&\|f\|_{\mathfrak{L}^{1}_\alpha(\mathcal{F}_{\mathcal{H}}})=
      \sum_{n=0}^\infty \alpha^n \mathrm{Tr}_{1,\ldots,n}|f_{n}(1,\ldots,n)|,
\end{eqnarray*}
where $\alpha>1$ is a real number. The subspace of finite sequences of degenerate operators whose kernels are infinitely
differentiable functions with compact supports will be denoted by $\mathfrak{L}^{1}_0(\mathcal{F}_{\mathcal{H}})$.

In the space $\mathfrak{L}_\gamma(\mathcal{F}_{\mathcal{H}})$ for arbitrary values of $t\in\mathbb{R}$, a one-parameter
family of mappings $\mathcal{G}(t)=\oplus_{n=0}^\infty\mathcal{G}_n(t)$ is defined, where the mapping $\mathcal{G}_n(t)$
in the subspace $\mathfrak{L}(\mathcal{H}_{n})$ is defined by the following formula
\begin{eqnarray}\label{grG}
&&\mathbb{R}^1\ni t\mapsto\mathcal{G}_n(t)b_n\doteq e^{itH_{n}}b_n e^{-itH_{n}},
\end{eqnarray}
and the self-adjoint operator
$H_{n}={\sum\limits}_{j=1}^{n}K(j)+{\sum\limits}_{j_{1}<j_{2}=1}^{n}\Phi(j_{1},j_{2})$ is the Hamiltonian of the system
of $n$ particles, i.e., the operator $K(j)$ is the kinetic energy operator of $j$ particle, $\Phi$ is the bounded pair
interaction potential operator.
In the space $\mathfrak{L}(\mathcal{H}_{n})$ the one-parameter mapping (\ref{grG}) forms a $\ast$-weakly continuous group
of operators, the infinitesimal generator of which, on the domain of its definition, coincides with the Heisenberg operator
\begin{eqnarray}\label{info}
    &&\mathcal{N}_n b_n\doteq-i\,(b_n H_n - H_n b_n).
\end{eqnarray}

The evolution of observables for many quantum particles, i.e., the sequences of operators
$O_n(t)=\mathcal{G}_n(t)O_n^0,\,n\geq1$, is governed by the Cauchy problem for the sequence of the Heisenberg
equations with these generators \cite{DL},\cite{Ger12}:
\begin{eqnarray}\label{Le}
   &&\hskip-7mm \frac{\partial}{\partial t}O_n(t)=\mathcal{N}_n O_n(t), \\
   &&\hskip-7mm  O_n(t)\big|_{t=0}=O_n^0,\quad n\geq1.\nonumber
\end{eqnarray}

For initial data $O_n^0\in \mathfrak{L}(\mathcal{H}_{n}),\, n\geq1$ of finite sequences of bounded operators with
infinitely differentiable kernels with compact supports the sequence $O_n(t)=\mathcal{G}_n(t)O_n^0,\,n\geq1,$
is a unique global-in-time classical solution to the Cauchy problem for the Heisenberg equations \eqref{Le}, and
for arbitrary initial data from $\mathfrak{L}(\mathcal{H}_{n}),\,n\geq1,$ is a unique global-in-time generalized
solution.

A similar result also holds for sequences of unbounded operators \cite{DL}.

The functional of the mean values (mathematical expectation) of sequences from the space
$\mathfrak{L}_\gamma(\mathcal{F}_{\mathcal{H}})$ is defined by a continuous linear functional, represented
by the following series expansion \cite{Ger12}:
\begin{eqnarray}\label{averageD}
     &&\langle b\rangle=(I,f)^{-1}(b,f)\doteq(I,f)^{-1}\sum\limits_{n=0}^{\infty}\frac{1}{n!}
         \,\mathrm{Tr}_{1,\ldots,n}\,b_{n}\,f_{n},
\end{eqnarray}
where the sequence $f\in \mathfrak{L}^1_\alpha(\mathcal{F}_{\mathcal{H}})$ and the factor
$(I,f)={\sum\limits}_{n=0}^{\infty}\frac{1}{n!}\mathrm{Tr}_{1,\ldots,n}f_{n}$ is the normalizing coefficient.
If $b\in\mathfrak{L}_\gamma(\mathcal{F}_{\mathcal{H}})$ and $f\in\mathfrak{L}^1_\alpha(\mathcal{F}_{\mathcal{H}})$
the functional \eqref{averageD} exists.

In the space $\mathfrak{L}^1_\alpha(\mathcal{F}_{\mathcal{H}})$, the dual in the sense of functional (\ref{averageD})
one-parameter mapping $\mathcal{G}^\ast(t)=\oplus_{n=0}^\infty \mathcal{G}^\ast_n(t)$, conjugate to group (\ref{grG}),
is defined as
\begin{eqnarray}\label{grGs}
    &&\mathbb{R}^1\ni t\mapsto\mathcal{G}^{\ast}_n(t)f_n\doteq e^{-itH_{n}}f_n e^{itH_{n}},
\end{eqnarray}
and forms a strongly continuous isometric group of operators that preserves positivity and self-adjointness
of operators \cite{DL},\cite{AF},\cite{DF}. The infinitesimal generator of the group (\ref{grGs}) on the subspace
$\mathfrak{L}^{1}_0(\mathcal{F}_{\mathcal{H}})$ coincides with the von Neumann operator
\begin{eqnarray}\label{infd}
    &&\mathcal{N}^{\ast}_n f_n\doteq-i\,(H_n f_n - f_n H_n).
\end{eqnarray}

The evolution of the state, i.e., the sequence of density operators $D_n(t)=\mathcal{G}^{\ast}_n(t)D_n^0,\,\,n\geq 1$,
(the kernel of a density operator is known as the density matrix) describes by the Cauchy problem for the sequence
of the von Neumann evolution equations for many quantum particles with generator (\ref{infd}):
\begin{eqnarray}\label{dLe}
   &&\hskip-7mm \frac{\partial}{\partial t}D_n(t)=\mathcal{N}^{\ast}_n D_n(t), \\
   &&\hskip-7mm D_n(t)\big|_{t=0}=D_n^0,\quad n\geq 1.\nonumber
\end{eqnarray}

For initial data $D_n^0\in \mathfrak{L}^{1}(\mathcal{H}_{n}),\,n\geq 1$ of finite sequences of operators with
infinitely differentiable kernels with compact supports, the sequence $D_n(t)=\mathcal{G}^{\ast}_n(t)D_n^0,\,n\geq 1$,
is a unique global-in-time strong solution, whereas for arbitrary initial data from the spaces
$\mathfrak{L}^{1}(\mathcal{H}_{n}),\,n\geq 1$, it is a unique global-in-time weak solution.

Thus, the evolution of finitely many quantum particles is governed by the fundamental evolution
equations, such as the Heisenberg equation for observables \eqref{Le} or its dual equation for a
state \eqref{dLe}.

\textcolor{blue!55!black}{\subsection{Cluster and cumulant expansion of operator groups}}

In the article, we primarily examine the hierarchies of evolution equations that describe how a system's state,
a composite of many quantum particles, evolves, and we provide a classification of these equations. To achieve
this, we introduce the concept of cumulants (semi-invariants \cite{K62}) of operator groups (\ref{grGs}), which
represent their connected parts.

The cumulant of the $n$th order of operator groups (\ref{grGs}) is defined by the following expansions \cite{Ger12}:
\begin{equation}\label{cum}
   \mathfrak{A}^\ast_{n}(t,1,\ldots,n)\doteq
      \sum\limits_{\scriptsize\mathrm{P}:\,(1,\ldots,n)={\bigcup}_i X_i}(-1)^{|\mathrm{P}|-1}(|\mathrm{P}|-1)!\,
       \prod_{X_i\subset \mathrm{P}}\mathcal{G}^\ast_{|X_i|}(t,X_i),
\end{equation}
where the symbol $\sum_\mathrm{P}$ denotes the sum over all partitions of $\mathrm{P}$ of the set of indices
$(1,\ldots,n)$ into $|\mathrm{P}|$ nonempty subsets $X_i\in(1,\ldots,n)$ that do not intersect. Let us give
examples of cumulant expansions \eqref{cum} of operator groups (\ref{grGs}):
\begin{eqnarray*}
    &&\mathfrak{A}^\ast_{1}(t,1)=\mathcal{G}^\ast_{1}(t,1), \\
    &&\mathfrak{A}^\ast_{2}(t,1,2)=\mathcal{G}^\ast_{2}(t,1,2)-\mathcal{G}^\ast_{1}(t,1)\mathcal{G}^\ast_{1}(t,2),\\
    &&\mathfrak{A}^\ast_{3}(t,1,2,3)=\mathcal{G}^\ast_{3}(t,1,2,3)-\mathcal{G}^\ast_{1}(t,1)\mathcal{G}^\ast_{2}(t,2,3)-
    \mathcal{G}^\ast_{1}(t,2)\mathcal{G}^\ast_{2}(t,1,3)-\\
    &&\hskip+7mm \mathcal{G}^\ast_{1}(t,3)\mathcal{G}^\ast_{2}(t,1,2)+
    2!\,\mathcal{G}^\ast_{1}(t,1)\mathcal{G}^\ast_{1}(t,2)\mathcal{G}^\ast_{1}(t,3),\\
   && \vdots
\end{eqnarray*}
Note that the cumulants of operator groups \eqref{cum} are solutions of recursive equations known as cluster
expansions of operator groups (\ref{grGs})

\begin{equation}\label{exp}
    \mathcal{G}^\ast_{n}(t,1,\ldots,n)=
      \sum\limits_{\scriptsize\mathrm{P}:\,(1,\ldots,n)=
      {\bigcup}_i X_i}\prod_{X_i\subset \mathrm{P}}\mathfrak{A}^\ast_{|X_i|}(t,X_i),\quad n\geq 1,
\end{equation}
where the notation of the formula \eqref{cum} is used. The structure of cluster expansions is determined by the structure
of the infinitesimal generator \eqref{infd} of operator groups (\ref{grGs}).

Let us present some characteristic properties of cumulants of operator groups \eqref{cum}.

In the case of a quantum system of non-interacting particles, we have
\begin{eqnarray*}
   &&\mathfrak{A}^\ast_{n}(t,1,\ldots,n)=0,\quad n\geq2.
\end{eqnarray*}
Indeed, since in this case the equality $\mathcal{G}^\ast_{n}(t,1,\ldots,n)=\prod_{j=1}^{n}\mathcal{G}^\ast_{1}(t,j),$ holds
for operator groups (\ref{grGs}), we establish
\begin{eqnarray*}
  &&\mathfrak{A}^\ast_{n}(t,1,\ldots,n)=\sum\limits_{\scriptsize\mathrm{P}:\,(1,\ldots,n)={\bigcup}_i X_i}
     (-1)^{|\mathrm{P}|-1}(|\mathrm{P}|-1)!\prod\limits_{X_{i}\subset
     \mathrm{P}}\prod_{j_i=1}^{|X_{i}|}\mathcal{G}^\ast_{1}(t,j_i)=\\
  &&=\sum\limits_{k=1}^{n}(-1)^{k-1}\mathrm{s}(n,k)(k-1)!\prod_{j=1}^{n}\mathcal{G}_{1}(t,j)=0,
\end{eqnarray*}
where $\mathrm{s}(n,k)$ are the Stirling numbers of the second kind, and the equality is used
\begin{eqnarray*}\label{Stirl}
  &&\hskip-5mm \sum\limits_{\mathrm{P}:\,(1,\ldots,n)={\bigcup}_i X_i}
    (-1)^{| \mathrm{P}|-1}(|\mathrm{P}|-1)!=
    \sum\limits_{k=1}^{n}(-1)^{k-1}\mathrm{s}(n,k)(k-1)!=\delta_{n,1},
\end{eqnarray*}
and $\delta_{n,1}$ is the Kronecker symbol.

If $f_{n}\in\mathfrak{L}^1(\mathcal{H}_{n})$, then for the cumulant of $n$th order \eqref{cum},
the following estimate is valid
\begin{eqnarray}\label{est}
   &&\big\|\mathfrak{A}^\ast_{n}(t,1,\ldots,n)f_{n}\big\|_{\mathfrak{L}^1(\mathcal{H}_{n})}
      \leq\sum\limits_{\scriptsize\mathrm{P}:\,(1,\ldots,n)={\bigcup}_i X_i}
      (|\mathrm{P}|-1)!\,\big\|f_{n}\big\|_{\mathfrak{L}^1(\mathcal{H}_{n})}\leq \\
   &&\leq\sum\limits_{k=1}^{n}\mathrm{s}(n,k)(k-1)!\big\|f_{n}\big\|_{\mathfrak{L}^1(\mathcal{H}_{n})}
      \leq n!\,e^{n}\,\big\|f_{n}\big\|_{\mathfrak{L}^1(\mathcal{H}_{n})},\nonumber
\end{eqnarray}
where $\mathrm{s}(n,k)$ are the Stirling numbers of the second kind.

We will also introduce some generalizations of the concept of cumulant expansions of operator groups for
clusters of quantum particles.

In what follows, we will use such abridged notations.
The symbol $(\{i_1,\ldots,i_s\})$ denotes a connected subset of indices $(i_1,\ldots,i_s)$ of the set of indices
$(1,\ldots,n)$, i.e., a set consisting of a single index element. Then the set
$(\{X_{1}\},\ldots,\{X_{|\mathrm{P}|}\})\in(1,\ldots,n)$ is the set of the $|\mathrm{P}|$ elements that are
mutually disjoint subsets $X_{i}\subset (1,\ldots,n)$ of the partition $\mathrm{P}$ of the set of indices
$(1,\ldots,n)={\bigcup\limits}_{i=1}^{|\mathrm{P}|}X_i.$

The cumulant of the $|\mathrm{P}|$th order of operator groups (\ref{grGs}) for $|\mathrm{P}|$ clusters of $n$ quantum
particles is denoted by $\mathfrak{A}^{\ast}_{|\mathrm{P}|}(t,\{X_{1}\},\ldots,\{X_{|\mathrm{P}|}\})$, and it is
defined by the expansion
\begin{eqnarray} \label{cumulantP}
   &&\hskip-8mm \mathfrak{A}^{\ast}_{|\mathrm{P}|}(t,\{X_1\},\ldots,\{X_{|\mathrm{P}|}\})\doteq\\
   &&\sum\limits_{\mathrm{P}^{'}:\,(\{X_1\},\ldots,\{X_{|\mathrm{P}|}\})=
      \bigcup_k Y_k}(-1)^{|\mathrm{P}^{'}|-1}({|\mathrm{P}^{'}|-1})!
      \prod\limits_{Y_k\subset\mathrm{P}^{'}}\mathcal{G}^{\ast}_{|\theta(Y_{k})|}(t,\theta(Y_{k})),\nonumber
\end{eqnarray}
where the notations are similar to that in formula \eqref{cum} and the operator $\theta$ is the declusterization
mapping defined by $\theta(\{X_1\},\ldots,\{X_{|\mathrm{P}|}\})\doteq(1,\ldots,n)$.

Note that the cumulant of the $|\mathrm{P}|$th order of operator groups (\ref{cumulantP}) for the $|\mathrm{P}|$
clusters of $n$ quantum particles is expressed in terms of the cumulants of the operator groups (\ref{cum}) of
$n$ particles. A typical example of the cumulant of a cluster of $s$ particles and $n$ particles is determined
by the following expansions:
\begin{eqnarray*}
    &&\hskip-8mm \mathfrak{A}^{\ast}_{1+n}(t,\{1,\ldots,s\},s+1,\ldots,s+n)=\\
    &&\sum\limits_{\scriptsize\mathrm{P}:(\{1,\ldots,s\},s+1,\ldots,s+n)=\bigcup_i X_i}
       (-1)^{|\mathrm{P}|-1}(|\mathrm{P}|-1)!\times \\
    &&\prod_{X_i\subset\mathrm{P}}\,\sum\limits_{\mathrm{P'}:\,\theta(X_{i})=\bigcup_{j_i} Y_{j_i}}\,
      \prod_{Y_{j_i}\subset\mathrm{P'}}\mathfrak{A}^{\ast}_{|Y_{j_i}|}(t,Y_{j_i}), \quad  n\geq0,
\end{eqnarray*}
for example,
\begin{eqnarray*}
    &&\mathfrak{A}^{\ast}_{1+0}(t,\{1,2\})=\mathfrak{A}^{\ast}_{2}(t,1,2)+
      \mathfrak{A}^{\ast}_{1}(t,1)\mathfrak{A}^{\ast}_{1}(t,2),\\
    &&\mathfrak{A}^{\ast}_{1+1}(t,\{1,2\},3)=\mathfrak{A}^{\ast}_{3}(t,1,2,3)+
      \mathfrak{A}^{\ast}_{2}(t,1,3)\mathfrak{A}^{\ast}_{1}(t,2)+
      \mathfrak{A}^{\ast}_{2}(t,2,3)\mathfrak{A}^{\ast}_{1}(t,1),\\
   && \vdots
\end{eqnarray*}
The cumulants of operator groups for clusters of quantum particles are the generating operators of the
expansions for non-perturbative solutions of hierarchies of evolution equations.

\textcolor{blue!55!black}{\subsection{Group of nonlinear operators of the von Neumann hierarchy}}

An alternative way to describe the states of quantum particle systems involves using operators defined as
cumulant expansions of density operators, which are interpreted as correlations of particle states. In this
subsection, we study a one-parameter group of nonlinear operators that defines a non-perturbative representation
of the solution to the von Neumann hierarchy for correlation operators.

For $f\in\mathfrak{L}^1_\alpha(\mathcal{F}_{\mathcal{H}})$, the following one-parameter family of nonlinear
mappings is defined
\begin{eqnarray}\label{groupKum}
   &&\mathbb{R}^1\ni t\mapsto \mathcal{G}(t;1,\ldots,n\mid f)\doteq \\
   &&\sum\limits_{\scriptsize\mathrm{P}:\,(1,\ldots,n)={\bigcup}_i X_i}
        \mathfrak{A}^{\ast}_{|\mathrm{P}|}(t,\{X_1\},\ldots,\{X_{|\mathrm{P}|}\})
        \mathop{\otimes}_{X_i\subset \mathrm{P}}f_{|{X_i}|}(X_i),\nonumber
\end{eqnarray}
where generating operators are cumulants (\ref{cumulantP}) of the corresponding order of groups
of operators (\ref{grGs}).
The following estimate is true
\begin{eqnarray*}
   &&\big\|\mathcal{G}(t;1,\ldots,n\mid f)\big\|_{\mathfrak{L}^{1}(\mathcal{H}_{n})}\leq n!e^{2n}c^{n},
\end{eqnarray*}
where
$c\equiv e^{3}\max(1,\max_{\mathrm{P}:\,(1,\ldots,n)=\bigcup_iX_i}\|f_{|X_{i}|}\|_{\mathfrak{L}^{1}(\mathcal{H}_{|X_{i}|})})$.

The properties of the one-parameter mapping (\ref{groupKum}) are described in the following theorem.

\bigskip
\begin{theorem}
If $f_{n}\in\mathfrak{L}^{1}(\mathcal{H}_{n}),\,n\geq1,$ then the one-parameter family of nonlinear operators
(\ref{groupKum}) is a bounded strong continuous group of nonlinear operators.
On the subspace $\mathfrak{L}^{1}_{0}(\mathcal{H}_{n})\subset\mathfrak{L}^{1}(\mathcal{H}_{n})$ the infinitesimal
generator of the group of nonlinear operators \eqref{groupKum} is defined by the following operator
\begin{eqnarray}\label{Nnl}
        &&\mathcal{N}(1,\ldots,n\mid f)\doteq\mathcal{N}^{\ast}_{n}(1,\ldots,n)f_{n}(1,\ldots,n)+\\
        &&\sum\limits_{\mathrm{P}:\,(1,\ldots,n)=X_{1}\bigcup X_2}\,\sum\limits_{i_{1}\in X_{1}}
           \sum\limits_{i_{2}\in X_{2}}\mathcal{N}_{\mathrm{int}}^{\ast}(i_{1},i_{2})
           f_{|X_{1}|}(t,X_{1})\otimes f_{|X_{2}|}(t,X_{2}), \nonumber
\end{eqnarray}
where the symbol ${\sum\limits}_{\mathrm{P}:\,(1,\ldots,n)=X_{1}\bigcup X_2}$ means the sum over all possible
partitions $\mathrm{P}$ of the set $(1,\ldots,n)$ into two nonempty mutually disjoint subsets $X_1$ and $X_2$
and the operator $\mathcal{N}^{\ast}_{\mathrm{int}}$ is defined by the formula
$\mathcal{N}^{\ast}_{\mathrm{int}}(i_{1},i_{2})f_n\doteq-i\,(\Phi(i_{1},i_{2})f_n-f_n\Phi(i_{1},i_{2}))$.
\end{theorem}

The scheme of the proof of this theorem is analogous to the proof of the existence theorem for the von Neumann
equations (\ref{dLe}) in the space of sequences of trace-class operators \cite{DL},\cite{Ger12}.

The evolution of all possible states of a quantum system of non-fixed, i.e. arbitrary but finite,
number of identical particles, obeying the Maxwell--Boltzmann statistics, is described by means of the
sequence $g(t)=(g_0,g_{1}(t),\ldots,g_{n}(t),\ldots)\in\mathfrak{L}^{1}(\mathcal{F}_\mathcal{H})$
of the correlation operators $g_{n}(t)=g_{n}(t,1,\ldots,n),\,n\geq1$, governed by the Cauchy problem
of the von Neumann hierarchy \cite{Ger12}:
\begin{eqnarray}\label{vNh}
   &&\hskip-5mm\frac{\partial}{\partial t}g_{n}(t,1,\ldots,n)=\mathcal{N}^{\ast}_{n}(1,\ldots,n)g_{n}(t,1,\ldots,n)+\\
   &&\sum_{\scriptsize\mathrm{P}:\,(1,\ldots,n)=X_{1}\bigcup X_2}\,\sum\limits_{i_{1}\in X_{1}}
      \sum\limits_{i_{2}\in X_{2}}\mathcal{N}_{\mathrm{int}}^{\ast}(i_{1},i_{2})
      g_{|X_{1}|}(t,X_{1})\otimes g_{|X_{2}|}(t,X_{2}), \nonumber\\
 \label{vNhi}
   &&\hskip-5mm g_{n}(t)\big|_{t=0}=g_{n}^{0},\quad n\geq1,
\end{eqnarray}
where the symbol ${\sum\limits}_{\mathrm{P}:\,(1,\ldots,n)=X_{1}\bigcup X_2}$ means the sum over all possible
partitions $\mathrm{P}$ of the set $(1,\ldots,n)$ into two nonempty mutually disjoint subsets $X_1$ and $X_2$,
as was accepted above.

The non-perturbative solution of the Cauchy problem of the von Neumann hierarchy (\ref{vNh}),(\ref{vNhi}) is
represented by the group of nonlinear operators (\ref{groupKum})
\begin{eqnarray}\label{ghs}
   &&g(t,1,\ldots,n)=\mathcal{G}(t;1,\ldots,n\mid g(0)), \quad n\geq1,
\end{eqnarray}
where the one-parameter mapping is defined by the formula (\ref{groupKum}), and the sequence of initial correlation
operators (\ref{vNhi}) is denoted as $g(0)=(g_0,g_{1}^{0},\ldots,g_{n}^{0},\ldots)$, and $g_0\in\mathbb{C}$. The
simplest examples of correlation operators (\ref{ghs}) are represented by such expansions:
\begin{eqnarray*}
   && g_{1}(t,1)=\mathfrak{A}^\ast_{1}(t,1)g_{1}^{0}(1),\\
   && g_{2}(t,1,2)=\mathfrak{A}^\ast_{1}(t,\{1,2\})g_{2}^{0}(1,2)+
     \mathfrak{A}^\ast_{2}(t,1,2)g_{1}^{0}(1)\otimes g_{1}^{0}(2),\\
   &&\vdots
\end{eqnarray*}

The following statement is true \cite{Ger12}.

In the case of bounded interaction potentials, the solution to the Cauchy problem of the von Neumann
hierarchy (\ref{vNh}),(\ref{vNhi}) is determined by the sequence of correlation operators represented
by expansions (\ref{ghs}).
If $g_{n}^{0}\in\mathfrak{L}^{1}_{0}(\mathcal{H}_{n})\subset\mathfrak{L}^{1}(\mathcal{H}_{n}),\,n\geq 1$,
then this is a global-in-time strong solution, and for arbitrary initial data
$g_{n}^{0}\in\mathfrak{L}^{1}(\mathcal{H}_{n}),\,n\geq 1$, it is a global-in-time weak solution.

This statement also holds for interaction potentials that satisfy the Kato conditions \cite{DL}.

In particular, for separable (product) initial states, the sequence of initial correlation operators takes
the form $g(0)=(0,g_{1}^{0},0,\ldots,0,\ldots)$. Then solution (\ref{ghs}) of the Cauchy problem of the
von Neumann hierarchy (\ref{vNh}),(\ref{vNhi}) is represented as follows:
\begin{eqnarray*}
   &&\hskip-7mm g_{n}(t,1,\ldots,n)=\mathfrak{A}^{\ast}_{n}(t,1,\ldots,n)\,\mathop{\otimes}_{i=1}^{n}g_{1}^{0}(i),
   \quad n\geq1,
\end{eqnarray*}
where the generating operator $\mathfrak{A}^{\ast}_{n}(t)$ is the $n$th-order cumulant of operator groups
(\ref{grG}) determined by expansion \eqref{cum}.

Note that the non-perturbative solution (\ref{ghs}) of the Cauchy problem of the von Neumann hierarchy
(\ref{vNh}),(\ref{vNhi}) is reduced to an iterative (perturbative) series as a result of applying analogues
of the Duhamel equations for cumulants (\ref{cumulantP}) of operator groups.

Thus, the cumulant origin of the operators for state correlations determines a cumulant structure of
the generating operators for expansions (\ref{groupKum}) that describe the propagation of correlations
(\ref{ghs}) in systems of many quantum particles.

\textcolor{blue!55!black}{\section{Non-perturbative solutions of hierarchies of evolution equations for quantum systems}}\label{sec3}

Hierarchies of evolution equations for sequences of reduced operators of a state for a finite number
of quantum particles are equivalent to the von Neumann equations. Their advantages consist in the
possibility of rigorously describing the evolution of infinitely many particles whose collective
behavior is characterized by statistical features, namely, the existence of an equilibrium state,
as well as the kinetic or hydrodynamic behavior in appropriate scaling approximations.

\textcolor{blue!55!black}{\subsection{Group of operators of the BBGKY hierarchy}}

Traditionally, the evolution of states of quantum many-particle systems is described by a solution
to the BBGKY hierarchy constructed using perturbation theory methods. In this subsection, we study a
one-parameter group of operators that defines a non-perturbative representation of the solution to
the BBGKY hierarchy.

The following one-parameter family of mapping is defined on the space $\mathfrak{L}^1_\alpha(\mathcal{F}_{\mathcal{H}})$
\begin{eqnarray}\label{RozvBBGKY}
   &&\hskip-7mm \mathbb{R}^1\ni t\mapsto(U^{\ast}(t)f)_{s}(1,\ldots,s)\doteq\\
   &&\sum\limits_{n=0}^{\infty}\frac{1}{n!}\,\mathrm{Tr}_{\mathrm{s+1,\ldots,s+n}}
      \mathfrak{A}^{\ast}_{1+n}(t,\{1,\ldots,s\},s+1,\ldots,s+n)f_{s+n}(1,\ldots,s+n),\nonumber
\end{eqnarray}
where the generating operators of the series expansion are the cumulants (\ref{cumulantP}) of the corresponding
order of the operator groups (\ref{grGs}) for the particle cluster and particles. The series representing mapping
(\ref{RozvBBGKY}) converges, and as a consequence of inequality (\ref{est}), the following estimate holds:
\begin{eqnarray*}
    &&\hskip-7mm \|U^{\ast}(t)f\|_{\mathfrak{L}^{1}_{\alpha}(\mathcal{F}_{\mathcal{H}})}\leq
        e^{2}(1-\frac{e}{\alpha})^{-1}\|f\|_{\mathfrak{L}^{1}_{\alpha}(\mathcal{F}_{\mathcal{H}})}.
\end{eqnarray*}

The properties of the one-parameter mapping (\ref{RozvBBGKY}) are described in the following theorem.

\bigskip
\begin{theorem}
If $f_s\in\mathfrak{L}^{1}(\mathcal{H}_s)$, then one-parameter mapping (\ref{RozvBBGKY}) is strongly
continuous group of bounded operators.
On the subspace $\mathfrak{L}^{1}_{0}(\mathcal{H}_s)\subset\mathfrak{L}^{1}(\mathcal{H}_s)$ the
infinitesimal generator of the operator groups \eqref{RozvBBGKY} is defined by the operator
\begin{eqnarray}\label{1}
        &&\hskip-7mm ((\mathcal{N}^\ast +[\mathfrak{a},\mathcal{N}^\ast])f)_{s}(1,\ldots,s)\doteq\\
        &&(\mathcal{N}^{\ast}f)_{s}(1,\ldots,s)+
            \sum_{j=1}^s\,\mathrm{Tr}_{s+1}\,\mathcal{N}^\ast_{\mathrm{int}}(j,s+1)f_{s+1}(1,\ldots,s,s+1),\nonumber
\end{eqnarray}
where the symbol $[\cdot,\cdot]$ denotes the commutator of the von Neumann operator
(\ref{infd}) which is the generator of the operator groups (\ref{grGs}) and the operator:
$(\mathfrak{a}f)_{s}\doteq\mathrm{Tr}_{s+1}\,f_{s+1},\, s\geq 1.$
\end{theorem}

The scheme of the proof of this theorem is analogous to the proof of the existence theorem for the von Neumann
equations (\ref{dLe}) in the space of sequences of trace-class operators \cite{DL},\cite{Ger12}.

The evolution of all possible states of a quantum system of non-fixed, i.e. arbitrary but finite,
number of identical particles, as an equivalent method compared with the density operators, is described
by means of the sequence $F(t)=(I,F_{1}(t),\ldots,F_{s}(t),\ldots)\in\mathfrak{L}^{1}(\mathcal{F}_\mathcal{H})$
of the reduced density operators $F_{s}(t)=F_{s}(t,1,\ldots,s),\,s\geq1$, governed by the Cauchy problem
of the BBGKY hierarchy \cite{Ger12}:
\begin{eqnarray}\label{h}
&&\hskip-5mm\frac{d}{dt}F(t)=\mathcal{N}^\ast F(t)+\big[\mathfrak{a},\mathcal{N}^\ast\big]F(t),\\
\label{hi}
&&\hskip-5mmF(t)|_{t=0}=F(0),
\end{eqnarray}
where the generator of this hierarchy of evolution equations is the operator (\ref{1}).
We remark that the structure of the generator of the BBGKY hierarchy of evolution equations for quantum
systems of particles with an arbitrary $n$-ary interaction potential \cite{Ger12},\cite{G09} has the form:
$e^\mathfrak{a}\mathcal{N}^\ast e^{-\mathfrak{a}}$. For the pairwise interaction potential, the following equality
holds: $e^\mathfrak{a}\mathcal{N}^\ast e^{-\mathfrak{a}}=\mathcal{N}^\ast+[\mathfrak{a},\mathcal{N}^\ast]$.

The non-perturbative solution of the Cauchy problem of the BBGKY hierarchy (\ref{h}),(\ref{hi}) is
represented by the operator group (\ref{RozvBBGKY})
\begin{eqnarray}\label{se}
   &&\hskip-5mm F(t)=U^{\ast}(t)F(0),
\end{eqnarray}
where the component-wise form of this expression is represented by expansions (\ref{RozvBBGKY}).

The following statement is true.

In the case of bounded interaction potentials, if $\alpha>e$, the solution to the Cauchy problem of the
BBGKY hierarchy (\ref{h}),(\ref{hi}) is determined by the sequence of reduced density operators
represented by expansions (\ref{se}).
If $F_{s}^{0}\in\mathfrak{L}^{1}_{0}(\mathcal{H}_{s})\subset\mathfrak{L}^{1}(\mathcal{H}_{s}),\,s\geq 1$,
then this is a global-in-time strong solution, and for arbitrary initial data
$F_{s}^{0}\in\mathfrak{L}^{1}(\mathcal{H}_{s}),\,s\geq 1$, it is a global-in-time weak solution.

Note that the definition of reduced density operators is formulated based on correlation operators
(\ref{ghs}) by means of the following series expansion \cite{GG22}:
\begin{eqnarray}\label{Fcluster}
    &&\hskip-7mm F_{s}(t,1,\ldots,s)\doteq \sum\limits_{n=0}^{\infty}\frac{1}{n!}\,
       \mathrm{Tr}_{\mathrm{s+1,\ldots,s+n}}\,g_{1+n}(t,\{1,\ldots,s\},s+1,\ldots,s+n), \,\, s\geq1,
\end{eqnarray}
where the correlation operators of the cluster of $s$ particles and particles $g_{1+n}(t), n\geq0,$ are
determined by expansions (\ref{ghs}). The redefinition of reduced density operators (\ref{Fcluster})
naturally arises from their definition within the grand canonical ensemble when explicitly taking into
account the normalization factor \cite{GG22}.

Thus, the cumulant origin of the operators for state correlations determines a cumulant structure of the
generating operators for expansions (\ref{se}) that describe the state evolution both for a finite average
number of quantum particles and the dynamics of an infinite number of quantum particles.

The cluster expansions (\ref{exp}) of operator groups (\ref{grGs}) underlie the classification of possible
solution representations \cite{GG23d} for the Cauchy problem of the BBGKY hierarchy (\ref{h}),(\ref{hi}).
In particular, the non-perturbative solution (\ref{se}) of the BBGKY hierarchy for many quantum particles
can be expressed as a perturbation (iteration) series by applying analogs of the Duhamel equation to the
cumulants (\ref{cumulantP}) of operator groups. Traditionally, perturbation theory methods are used to
study the rigorous derivation of quantum kinetic equations in the scaling limits of the solution of the
BBGKY hierarchy.

\textcolor{blue!55!black}{\subsection{Hierarchy of nonlinear evolution equations for reduced correlation
operators}}\label{subsec2}

An equivalent method for describing the states of quantum many-particle systems consists of using operators
defined by the cluster expansions of reduced density operators, which are interpreted as reduced correlation
operators. On a microscopic scale, the macroscopic characteristics of fluctuations of observables are directly
determined by these reduced correlation operators \cite{BogLect},\cite{G17}.

The following one-parameter family of mapping is defined on the space $\mathfrak{L}^1_\alpha(\mathcal{F}_{\mathcal{H}})$
\begin{eqnarray}\label{Rozvc}
   &&\hskip-5mm \mathbb{R}^1\ni t\mapsto U(t;1,\ldots,s\mid f)\doteq\\
   &&\sum\limits_{n=0}^{\infty}\frac{1}{n!}
        \,\mathrm{Tr}_{s+1,\ldots,s+n}\,\mathfrak{A}_{1+n}(t;\{1,\ldots,s\},s+1,\ldots,s+n\mid f),\nonumber
\end{eqnarray}
where the generating operator $\mathfrak{A}_{1+n}(t\mid f)$ of this series is the $(1+n)$th-order cumulant
of groups of nonlinear operators (\ref{groupKum}):
\begin{eqnarray}\label{cc}
   &&\hskip-7mm\mathfrak{A}_{1+n}(t;\{1,\ldots,s\},s+1,\ldots,s+n\mid f)\doteq\\
   &&\sum\limits_{\mathrm{P}:\,(\{1,\ldots,s\},s+1,\ldots,s+n)=
      \bigcup_k X_k}(-1)^{|\mathrm{P}|-1}({|\mathrm{P}|-1})!\,
      \mathcal{G}(t;\theta(X_1)\mid\ldots\nonumber\\
   &&\mathcal{G}(t;\theta(X_{|\mathrm{P}|})\mid f)\ldots), \quad n\geq0,\nonumber
\end{eqnarray}
and the composition of mappings (\ref{groupKum}) of the corresponding noninteracting clusters of particles
we denoted by $\mathcal{G}(t;\theta(X_1)\mid \ldots\mathcal{G}(t;\theta(X_{|\mathrm{P}|})\mid f)\ldots)$,
for example,
\begin{eqnarray*}
    &&\hskip-7mm\mathcal{G}\big(t;1\mid\mathcal{G}(t;2\mid f)\big)=
        \mathfrak{A}_{1}(t,1)\mathfrak{A}_{1}(t,2)f_{2}(1,2),\\
    &&\hskip-7mm\mathcal{G}\big(t;1,2\mid\mathcal{G}(t;3\mid f)\big)=
        \mathfrak{A}_{1}(t,\{1,2\})\mathfrak{A}_{1}(t,3)f_{3}(1,2,3)+\\
    &&\mathfrak{A}_{2}(t,1,2)\mathfrak{A}_{1}(t,3)\big(f_{1}(1)\otimes f_{2}(2,3)+f_{1}(2)\otimes f_{2}(1,3)\big).
\end{eqnarray*}

Below we adduce the examples of expansions (\ref{cc}). The first-order cumulant of the groups
of nonlinear operators (\ref{groupKum}) is the same group of nonlinear operators, i.e.,
\begin{eqnarray*}
     &&\mathfrak{A}_{1}(t;\{1,\ldots,s\}\mid f)=\mathcal{G}(t;1,\ldots,s\mid f).
\end{eqnarray*}
In case of $s=2$ the second-order cumulant of nonlinear operators (\ref{groupKum}) has the following structure:
\begin{eqnarray*}
     &&\hskip-7mm \mathfrak{A}_{1+1}(t;\{1,2\},3\mid f)=\mathcal{G}(t;1,2,3\mid f)-
       \mathcal{G}\big(t;1,2\mid\mathcal{G}(t;3\mid f)\big)=\\
     && \mathfrak{A}_{1+1}(t,\{1,2\},3)f_{3}(1,2,3)+\big(\mathfrak{A}_{1+1}(t,\{1,2\},3)-\\
     && \mathfrak{A}_{1+1}(t,2,3)\mathfrak{A}_{1}(t,1)\big)f_{1}(1)\otimes f_{2}(2,3)+\\
     &&\big(\mathfrak{A}_{1+1}(t,\{1,2\},3)-
        \mathfrak{A}_{1+1}(t,1,3)\mathfrak{A}_{1}(t,2)\big)f_{1}(2)\otimes f_{2}(1,3)+\\
     &&\mathfrak{A}_{1+1}(t,\{1,2\},3)f_{1}(3)\otimes f_{2}(1,2)+
        \mathfrak{A}_{3}(t,1,2,3)f_{1}(1)\otimes f_{1}(2)\otimes f_{1}(3),
\end{eqnarray*}
where the operator
\begin{equation*}
    \mathfrak{A}_{3}(t,1,2,3)=\mathfrak{A}_{1+1}(t,\{1,2\},3)-\mathfrak{A}_{1+1}(t,2,3)\mathfrak{A}_{1}(t,1)-
     \mathfrak{A}_{1+1}(t,1,3)\mathfrak{A}_{1}(t,2)
\end{equation*}
is the third-order cumulant (\ref{cum}) of operator groups (\ref{grGs}).

The series representing the mapping (\ref{Rozvc}) converges, and the following estimate holds:
\begin{eqnarray*}
    &&\hskip-7mm \big\|U(t;1,\ldots,s\mid f)\big\|_{\mathfrak{L}^{1}(\mathcal{H}_{s})}\leq
        s!(2e^2)^s\emph{c}^s\sum_{n=0}^{\infty}(2e^2)^n\emph{c}^n,
\end{eqnarray*}
where
$c\equiv e^{3}\max(1,\max_{\mathrm{P}:\,(1,\ldots,s)=\bigcup_iX_i}\|f_{|X_{i}|}\|_{\mathfrak{L}^{1}(\mathcal{H}_{|X_{i}|})})$.

The properties of the one-parameter mapping (\ref{RozvBBGKY}) are described in the following theorem.

\bigskip
\begin{theorem}
If $f\in\mathfrak{L}^1_\alpha(\mathcal{F}_{\mathcal{H}})$, then the one-parameter mapping (\ref{Rozvc}) is
a bounded strong continuous group of nonlinear operators.
On the subspace $\mathfrak{L}^{1}_{0}(\mathcal{F}_{\mathcal{H}})\subset\mathfrak{L}^1_\alpha(\mathcal{F}_{\mathcal{H}})$
the infinitesimal generator of the operator groups \eqref{Rozvc} is defined by the operator
\begin{eqnarray}\label{2}
        &&\hskip-5mm \big(\mathcal{N}(\cdot\mid f)+\mathfrak{a}\mathcal{N}(\cdot\mid f))-
           \mathcal{N}(\cdot\mid \mathfrak{a}f)\big)_{s}(1,\ldots,s)\doteq \\
        &&(\mathcal{N}^{\ast}_{s}f_{s})(1,\ldots,s)+\nonumber\\
        &&\sum\limits_{\mathrm{P}:\,(1,\ldots,s)=X_{1}\cup X_2}\,\sum\limits_{i_{1}\in X_{1}}
           \sum\limits_{i_{2}\in X_{2}}\mathcal{N}_{\mathrm{int}}^{\ast}(i_{1},i_{2})
           f_{|X_{1}|}(X_{1})\otimes f_{|X_{2}|}(X_{2}))+\nonumber\\
       &&\mathrm{Tr}_{s+1}\sum\limits_{j=1}^{s}\mathcal{N}^{\ast}_{\mathrm{int}}(i,s+1)\big(f_{s+1}(1,\ldots,s+1)+\nonumber\\
       &&\sum\limits_{\scriptsize\mathrm{P}:(1,\ldots,s+1)=X_1\cup X_2,\, i\in
             X_1;s+1\in X_2}f_{|X_1|}(X_1)\otimes f_{|X_2|}(X_2)\big),\nonumber
\end{eqnarray}
where we used the accepted notations for the infinitesimal generator \eqref{Nnl} of the group of nonlinear
operators \eqref{groupKum}.
\end{theorem}

The scheme of the proof of this theorem is analogous to the proof of the existence theorem for the von Neumann
hierarchy (\ref{vNh}) in the space of sequences of trace-class operators.

The evolution of all possible states of a quantum system, as an equivalent method compared with the reduced
density operators, is described by means of the sequence
$G(t)=(I,G_{1}(t),\ldots,G_{s}(t),\ldots)\in\mathfrak{L}^{1}(\mathcal{F}_\mathcal{H})$ of the reduced correlation
operators $G_{s}(t)=G_{s}(t,1,\ldots,s),\,s\geq1$, governed by the Cauchy problem of the following hierarchy of
nonlinear evolution equations \cite{Ger12}:
\begin{eqnarray}\label{ch}
     &&\hskip-5mm\frac{d}{dt}G(t)=\mathcal{N}(\cdot\mid G(t))+\mathfrak{a}\mathcal{N}(\cdot\mid G(t))
             -\mathcal{N}(\cdot\mid \mathfrak{a}G(t)),\\
      \label{chi}
     &&\hskip-5mm G(t)|_{t=0}=G(0),
\end{eqnarray}
where the generator of this hierarchy of evolution equations is the operator (\ref{2}).

The non-perturbative solution of the Cauchy problem of the hierarchy of nonlinear evolution equations
(\ref{ch}),(\ref{chi}) is represented by the group of nonlinear operators (\ref{Rozvc})
\begin{eqnarray}\label{cse}
   &&\hskip-5mm G(t)=U(t;\cdot\mid G(0)),
\end{eqnarray}
where the component-wise form of this expression is represented by expansions (\ref{Rozvc}).

The following statement is true \cite{G17},\cite{GP13}.

In the case of bounded interaction potentials, if $\alpha>e$, the solution to the Cauchy problem of hierarchy
(\ref{ch}),(\ref{chi}) is determined by the sequence of reduced correlation operators represented by expansions
(\ref{cse}).
If $G_{s}^{0}\in\mathfrak{L}^{1}_{0}(\mathcal{H}_{s})\subset\mathfrak{L}^{1}(\mathcal{H}_{s}),\,s\geq 1$, then
this is a global-in-time strong solution, and for arbitrary initial data
$G_{s}^{0}\in\mathfrak{L}^{1}(\mathcal{H}_{s}),\,s\geq 1$, it is a global-in-time weak solution.

In particular, for separable (product) initial states, the sequence of initial reduced correlation operators takes
the form $G(0)=(0,G_{1}^{0},0,\ldots,0,\ldots)$. Then solution (\ref{cse}) of the Cauchy problem (\ref{ch}),(\ref{chi})
is represented by the following series:
\begin{eqnarray}\label{mcc}
   &&\hskip-7mm G_{s}(t,1,\ldots,s)=\sum\limits_{n=0}^{\infty}\frac{1}{n!}
       \,\mathrm{Tr}_{s+1,\ldots,s+n}\,\mathfrak{A}^{\ast}_{s+n}(t;1,\ldots,s+n)
       \mathop{\otimes}_{i=1}^{s+n}G_1^{0}(i),\quad s\geq1,
\end{eqnarray}
where the generating operator $\mathfrak{A}^{\ast}_{s+n}(t)$ of series (\ref{mcc}) is the $(s+n)$th-order cumulant
(\ref{cum}) of operator groups (\ref{grG}).

It should be noted that the conventional definition of reduced correlation operators is formulated based
on cluster expansions of reduced density operators, namely
\begin{eqnarray*}
   &&F_{s}(t,1,\ldots,s)=
      \sum\limits_{\scriptsize\mathrm{P}:(1,\ldots,s)=\cup_{i}X_{i}}
      \mathop{\otimes}\limits_{X_i\subset\mathrm{P}}G_{|X_i|}(t,X_i),\quad s\geq1,
\end{eqnarray*}
where we used the above notations. Then, according to the definition of reduced density operators (\ref{Fcluster})
formulated based on correlation operators (\ref{ghs}), the reduced correlation operators represented by means of
the following series \cite{GG22}:
\begin{eqnarray}\label{Ccluster}
    &&\hskip-5mm G_{s}(t,1,\ldots,s)\doteq \sum\limits_{n=0}^{\infty}\frac{1}{n!}\,
       \mathrm{Tr}_{\mathrm{s+1,\ldots,s+n}}\,g_{s+n}(t,1,\ldots,s+n), \quad s\geq1,
\end{eqnarray}
where the correlation operator $g_{s+n}(t),$ of the $s+n$ particles is determined by expansion (\ref{ghs}).

Thus, the cumulant origin of the operators for state correlations (\ref{ghs}) determines a cumulant structure
of the generating operators for expansions (\ref{Ccluster}) that describe the evolution of correlations both
for a finite average number of quantum particles and the dynamics of an infinite number of quantum particles.

\textcolor{blue!55!black}{\section{Conclusion}}\label{sec4}

In the paper, it was established that the notion of cumulants (semi-invariants) of operator groups
underlies non-perturbative expansions of solutions for the fundamental evolution equations describing
the state evolution of quantum systems, namely, of the von Neumann hierarchy (\ref{vNh}) for correlation
operators, of the BBGKY hierarchy (\ref{h}) for reduced density operators, and of the hierarchy of
nonlinear evolution equations (\ref{ch}) for reduced correlation operators.
We note that for systems of collisional classical particles, the concept of cumulants of operator groups
is considered in the papers \cite{GG22},\cite{GG21}.

We emphasize that the structure of expansions for correlation operators, whose generating operators
are cumulants of the corresponding order of operator groups of a finite number of quantum particles,
induces the cumulant structure of series expansions for reduced density operators (\ref{se}), and reduced
correlation operators (\ref{cse}). Thus, in fact, the dynamics of states of quantum systems with an
infinite number of particles are substantially governed by the behavior of particle correlations.

As is well known, the BBGKY hierarchy is the dual hierarchy of evolution equations to the hierarchy
of evolution equations for observables \cite{BG},\cite{GG21}. In the Appendix, we demonstrate that
the concept of cumulants of operator groups of the Heisenberg equations also underlies non-perturbative
solutions of the hierarchy of evolution equations describing reduced observables of many quantum particles.

We emphasise that the work considered quantum systems of an unspecified number of identical spinless
particles, satisfying the Maxwell--Boltzmann statistics. The results formulated above can be extended
to systems of many bosons or fermions in accordance with the work \cite{GP}.
The article \cite{GG25} discusses the structure of expansions that represent non-perturbative solutions
of the Cauchy problem for the evolution equation hierarchies for the state and observables of many-particle
systems with topological nearest-neighbor interaction.

We note that the method for deriving kinetic equations from non-perturbative solutions of hierarchies
of evolution equations for the state of many-particle systems is based on the notion of kinetic
cluster expansions of cumulants of operator groups. In the works \cite{GT,G20,G26,GG}, the concept
of semi-invariants of cumulants of operator groups (\ref{cum}), which are solutions of the kinetic
cluster expansions for cumulants of operator groups, was developed.
Thus, the structure of non-perturbative solutions to hierarchies of fundamental evolution equations
relies on the concept of cumulants of operator groups. This concept also underlies the kinetic
description of the collective behavior of quantum systems with infinitely many particles.

\bigskip

\textbf{Acknowledgments.} \small{The author would like to thank Isaac Newton Institute and the London
Mathematical Society for their financial support during the visit to England, where this manuscript for
the conference "Evolution Equations, Operator Semigroups and Beyond" was prepared. Additionally, the
author appreciates the warm hospitality extended by the School of Mathematics and Physics at the
University of Surrey.}

\appendix

\textcolor{blue!55!black}{\section{Appendix: Cumulants of operator groups of the Heisenberg equations}}\label{secA1}

The motivation for describing the evolution of many-particle systems in terms of reduced observables
is related to possible equivalent representations of the mean value functional (mathematical expectation)
of observables. Specifically, the evolution of a system can be described in two ways: in terms of reduced
observables governed by a hierarchy of evolution equations, or in terms of a reduced density operators
governed by the hierarchy of dual evolution equations known as the BBGKY hierarchy. For quantum systems
with a finite number of particles, these hierarchies of evolution equations are equivalent to the Heisenberg
equation for observables and the von Neumann equation for density operator, respectively.

The following one-parameter family of mapping is defined in the space $\mathfrak{L}_{\gamma}(\mathcal{F}_\mathcal{H})$:
\begin{eqnarray}\label{krut_rozv}
       &&\hskip-7mm \mathbb{R}^{1}\ni t\mapsto\big(U(t)b\big) _{s}(1,\ldots,s)\doteq\\
       &&\sum_{n=0}^s\,\frac{1}{n!}\sum_{j_1\neq\ldots\neq j_{n}=1}^s
         \mathfrak{A}_{1+n}(t,\{(1,\ldots,s)\setminus(j_1,\ldots,j_{n})\},\nonumber\\
       &&j_1,\ldots,j_{n}\big)\,b_{s-n}((1,\ldots,s)\setminus(j_1,\ldots,j_{n}))\otimes
         \mathbbm{1}_{(j_1)}\ldots\otimes\mathbbm{1}_{(j_n)},\quad s\geq 1,\nonumber
\end{eqnarray}
where the generating operators of this expansion are the cumulants of the corresponding order
of the operator groups (\ref{grG}) for the particle cluster and particles.
By virtue of an analogue of inequality \eqref{est} for the expansion that defines the mapping
(\ref{krut_rozv}), the following estimate holds:
\begin{eqnarray*}
&&\|U(t)b\|_{\mathfrak{L}_{\gamma}(\mathcal{F}_{\mathcal{H}})}\leq
e^2\,(1-\gamma e)^{-1}\,\|b\|_{\mathfrak{L}_{\gamma}(\mathcal{F}_{\mathcal{H}})}.
\end{eqnarray*}

The properties of the one-parameter mapping (\ref{krut_rozv}) are described in the following theorem.

\medskip
\begin{theorem}
If $b\in\mathfrak{L}_{\gamma}(\mathcal{F}_\mathcal{H})$ and $\gamma<e^{-1}$, then the one-parameter
mapping (\ref{krut_rozv}) is a $C_{0}^{\ast}$-group.
The infinitesimal generator of this operator groups is a closed operator for the $\ast$-weak topology and
on the domain of the definition which is the everywhere dense set for the $\ast$-weak topology of the space
$\mathfrak{L}_{\gamma}(\mathcal{F}_\mathcal{H})$ it is defined by the operator
\begin{eqnarray}\label{d}
       &&\hskip-7mm ((\mathcal{N}+[\mathcal{N},\mathfrak{a}^+])b)_{s}(1,\ldots,s)\doteq (\mathcal{N}b)_{s}(1,\ldots,s)+\\
       &&\sum_{j_1\neq j_{2}=1}^s\mathcal{N}_{\mathrm{int}}(j_1,j_{2})b_{s-1}((1,\ldots,s)\setminus j_1)
            \otimes\mathbbm{1}_{(j_1)},\nonumber
\end{eqnarray}
where the Heisenberg operator $\mathcal{N}_{s}$ is the generator (\ref{info}) of the operator groups (\ref{grG}),
the symbol $[\cdot,\cdot]$ denotes the commutator of the operators $\mathcal{N}_{s}$ and the operator:
$(\mathfrak{a}^{+}b)_{s}(1,\ldots,s)\doteq\sum_{j=1}^s\,b_{s-1}((1,\ldots,s)\setminus j)\otimes\mathbbm{1}_{(j)},\,s\geq 1$.
\end{theorem}

\vskip-4mm
The proof of the theorem is analogous to the theorem for operator groups (\ref{grG}) of the Heisenberg equations
for many particles in the space of sequences of bounded operators \cite{DL},\cite{AF}.

As noted, an equivalent approach to describing the evolution of quantum many-particle systems is based on
sequences $B(t)=(B_0,B_{1}(t),\ldots,B_{s}(t),\ldots)\in\mathfrak{L}_{\gamma}(\mathcal{F}_\mathcal{H})$ of
the reduced observables $B_{s}(t)=B_{s}(t,1,\ldots,s),\,s\geq1$, which are defined by the Cauchy problem
to the hierarchy of evolution equations \cite{BG}:
\begin{eqnarray}\label{dh}
     &&\hskip-5mm\frac{d}{dt}B(t)=\mathcal{N}B(t)+\big[\mathcal{N},\mathfrak{a}^+\big]B(t),\\
  \label{dhi}
     &&\hskip-5mm B(t)|_{t=0}=B(0),
\end{eqnarray}
where the generator of this hierarchy of evolution equations is the operator (\ref{d}).
In component-wise form, the hierarchy of evolution equations for reduced observables \eqref{dh}
constitutes a system of recurrence evolution equations.

The non-perturbative solution of the Cauchy problem for the hierarchy of evolution equations (\ref{dh}),(\ref{dhi})
for a sequence of reduced observables is represented by the group of operators (\ref{krut_rozv})
\begin{eqnarray}\label{sed}
   &&\hskip-7mm B(t)=U(t)B(0).
\end{eqnarray}
We give examples of expansions (\ref{sed}) for reduced observables:
\begin{eqnarray*}
   &&B_{1}(t,1)=\mathfrak{A}_{1}(t,1)B_{1}^0(1),\\
   && B_{2}(t,1,2)=\mathfrak{A}_{1}(t,\{1,2\})B_{2}^0(1,2)+
      \mathfrak{A}_{2}(t,1,2)(B_{1}^0(1)\otimes\mathbbm{1}_{(2)}+\mathbbm{1}_{(1)}\otimes B_{1}^0(2)),\\
   &&\vdots
\end{eqnarray*}
It should also be noted that single-component sequences of reduced observables $B^{(1)}(0)=(0,b_{1}(1),0,\ldots)$
correspond to observables of additive type \cite{BG}. If the initial condition (\ref{dhi}) is specified by a sequence
of reduced observables of additive type, then the structure of the solution expansion (\ref{sed}) takes the form:
\begin{eqnarray*}
    &&\hskip-7mm B_{s}^{(1)}(t,1,\ldots,s)=\\
    &&\mathfrak{A}_{s}(t,1,\ldots,s)\sum_{j=1}^s \mathbbm{1}_{(1)}\otimes\ldots
        \otimes\mathbbm{1}_{(j-1)}\otimes b_{1}(j)\otimes\mathbbm{1}_{(j+1)}\otimes\ldots
        \otimes\mathbbm{1}_{(s)}, \quad s\geq 1,
\end{eqnarray*}
whose generating operator $\mathfrak{A}_{s}(t)$ is the $s$th-order cumulant (\ref{cum}) of operator groups (\ref{grG}).

In the space $\mathfrak{L}_{\gamma}(\mathcal{F}_{\mathcal{H}})$ the following existence theorem holds \cite{GG23d}.

The non-perturbative solution of the Cauchy problem (\ref{dh}),(\ref{dhi}) is represented by expansions (\ref{sed}),
in which the generating operators are the cumulants (\ref{cum}) of the operator groups (\ref{grG}) of the
corresponding order.
Provided the condition $\gamma<e^{-1}$ for the initial data $B(0)\in \mathfrak{L}_{0}(\mathcal{F}_{\mathcal{H}})$
the sequence (\ref{sed}) is the unique global-in-time classical solution, while for arbitrary initial data
$B(0)\in \mathfrak{L}_{\gamma}(\mathcal{F}_{\mathcal{H}})$, it is a unique global-in-time generalized solution.

The proof of this statement is analogous to the proof of the existence theorem for the Heisenberg equations
(\ref{Le}) for many quantum particles in the space of sequences of bounded operators \cite{Ger12}.
A similar outcome holds for sequences of unbounded operators that represent observables \cite{DL}.

\vskip+5mm
\small{
\addcontentsline{toc}{section}{References}


}

\end{document}